\documentclass[aps,preprint,amsmath,amssymb,11pt]{revtex4}
\usepackage{graphicx}
\usepackage{epstopdf}
\pdfoutput=1 
\usepackage[T1]{fontenc}
\newcommand\sss{\scriptscriptstyle}
\newcommand{\mW}{m_{\sss W}}
\newcommand{\sW}{s_{\sss W}}
\newcommand{\cW}{c_{\sss W}}

\begin{document}

\title{Prospective constraints on anomalous Higgs boson interactions in an effective Lagrangian via diphoton production at FCC-hh}
\author{H. Denizli}
\email[]{denizli_h@ibu.edu.tr}
\author{A. Senol}
\email[]{senol_a@ibu.edu.tr} 
\affiliation{Department of Physics, Bolu Abant Izzet Baysal University, 14280, Bolu, Turkey}
\begin{abstract}

We study the CP-conserving and CP-violating dimension-six operators of Higgs-gauge boson couplings via $pp\to\gamma\gamma$+n-jet signal process in a strongly interacting light Higgs based effective field theory framework at the center of mass energy of 100 TeV. In order to perform a simulation which includes realistic detector effects, the signal events in the existence of $\bar{c}_{\gamma} $, $\tilde{c}_{\gamma}$, $\bar{c}_{g} $ and $\tilde{c}_{g}$ Wilson coefficients and the relevant SM background events are generated in MadGraph, then passed through Pythia 8 for parton showering and finally run Delphes with FCC-hh detector card. In our analysis, we focus on the kinematic variables of the two photons in the final states of signal and relevant background processes that can reconstruct Higgs boson. We obtain constraints on the four  Wilson coefficients of dimension-six operators using the transverse momentum distribution of reconstructed di-photon system with optimized kinematic cuts. The obtained 95\% confidence level limits on these four Wilson coefficients including detector effects at $\sqrt s$=100 TeV with an integrated luminosity of 30 ab$^{-1}$ without systematic error are at least one order or more better than current experimental limits reported by ATLAS experiment. Even with $\delta_{sys}=2 \%$ systematic error, we find comparable limits with current experimental results.

\end{abstract}

\maketitle

\section{Introduction}

The particle physics has reached at a notable milestone in its history with the discovery of a scalar boson of 125 GeV in July 2012 at the LHC \cite{Aad:2012tfa,Chatrchyan:2012xdj}. The new discovered state by CMS and ATLAS experiments using collected proton-proton collision data at $\sqrt s = $7 TeV and 8 TeV was consistent with the characteristics of a Higgs boson which completes the matrix of particles and interactions in the Standard Model. However, not only experimental facts such as abundance of matter on antimatter, striking evidence of dark matter and non-zero neutrino masses, but also theoretical issues such as the problem of hierarchy, the dynamic origin of the Higgs mechanism requires the extension of the Standard Model. In addition, the existence of this new 125 GeV scalar field emerged with new challenging questions from phenomenological puzzles to riddles of deep quantum field theory. There are two ways to prove that SM is indeed a valid theory up to very high energy scales; i) the EW sector should be over-constrained and test the structure at the next leading order (NLO) corrections level or ii) there must be direct evidence for a possible dynamic explanation of the Higgs mechanism. Thus, the precise measurement of the Higgs boson properties will give us detailed information on the Electroweak Symmetry Breaking (EWSB) mechanism of the SM and new physics effects beyond the SM. In the literature, there have been many theoretical proposals to explain the origin of the EWSB such as the Higgs being elementary (as in the Standard Model) and weakly interacting \cite{Altarelli:2012dq} or being composite and related to a new strongly interacting sector \cite{Dimopoulos:1979es,Weinberg:1975gm}. However, latter one is exposed to strong constraints because of flavor changing neutral currents and precision electroweak measurements. Recent theoretical improvements provide opportunity the construction of models in agreement with the experimental bounds \cite{Hill:2002ap}. Further study of the Higgs boson couplings will play an important role in the searching for new situations related to the EWSB mechanism \cite{Buchmuller:1985jz,Grzadkowski:2010es}. One of the extremely useful tools for searches new physics in the Higgs sector is the Effective Field Theory (EFT) approach which has become very popular in the recent years \cite{Hagiwara:1993qt,Corbett:2012ja,Ellis:2014dva,Ellis:2014jta,Corbett:2015ksa,Aad:2015tna,Monfared:2016vwr,Englert:2015hrx,Englert:2016hvy,Degrande:2016dqg,Kilian:2017nio,Ellis:2017kfi,Ferreira:2016jea,Khanpour:2017inb,Denizli:2017pyu,Liu-Sheng:2017pxk,Khanpour:2017cfq,Kuday:2017vsh,Hesari:2018ssq,Kumar:2019bmk,Freitas:2019hbk,Li:2019evl,Denizli:2019ijf,Shi:2018lqf,Hays:2018zze,Aaboud:2018xdt}. In the EFT framework, the new physics associated with the EWSB effects on the phenomenology of the Higgs boson can be parametrized in terms of higher dimensional operators which are invariant under the SM symmetries and suppressed by the new physics scale $\Lambda$ as follows:
\begin{eqnarray}
\mathcal{L}_{EFT}=\mathcal{L}_{SM}+\sum_i\sum_{d > 4}\frac{c_d^{(i)}}{\Lambda^{d-4}}\mathcal{O}_d^{(i)}
\end{eqnarray}
where $d$ is the dimension of the operators, $c^{(i)}$ are the Wilson coefficients, $\mathcal {O}^{(i)}$ are all the gauge-invariant operators at mass-dimension $d$ involving the Standard Model fields, $\Lambda$ is a scale of new physics up to which the EFT is valid. The leading effects of new physics will be represented by the dimension-six operators is expected, since they are the least suppressed.

After the completion of the LHC and High-luminosity LHC physics programmes, the energy frontier collider project  having potential to search for wide parameter range of new physics are needed to precisely measure the Higgs self-coupling and fully explore the dynamic of EWSB on the TeV scale. The Future Circular Collider (FCC) Study is one of the future project currently under consideration by CERN which comes to fore with its unique 100 km tunnel infrastructure and technology as well as the physics opportunities \cite{Abada:2019lih}. This project covers synergy and complementarity of the three different colliders options; a luminosity-frontier highest-energy lepton collider (FCC-ee ) \cite{Abada:2019zxq}, an energy-frontier hadron collider (FCC-hh) \cite{Benedikt:2018csr} and a high energy hadron electron collider (FCC-he) \cite{Abada:2019lih}. The FCC-hh is designed to provide proton-proton collisions at the 100 TeV centre-of-mass energy with peak luminosity $5\times10^{34}$ cm$^{-2}$s$^{-1}$. Having this high center-of-mass-energy will increase cross sections for events in the partonic level which will than result in greater sensitivity to various interesting physics processes produced involving the Higgs bosons at high transverse momentum. Compared to other decay channels, the Higgs boson decay into two photons  is a particularly attractive opportunity to investigate the properties of the Higgs boson and to search for deviations from the Standard Model predictions due to beyond-Standard Model (BSM) processes. Despite  $H\to\gamma\gamma$ decay channel have small branching fraction of $\approx$ 0.2 \% \cite{Heinemeyer:2013tqa} predicted by the SM, it provides a clean final-state topology and a precise reconstruction of the diphoton mass. The dominant background arises from irreducible direct-diphoton production and from the reducible $pp\to\gamma\gamma$+jets and $pp\to$ jets final states.

In this study, we work out the effects of anomalous CP-even and CP-odd operators described with an EFT effective Lagrange between the Higgs boson and gluons as well as Higgs boson and photons via $pp\to\gamma\gamma$+n-jet process; di-photon production with up to two additional partons ($n=0,1,2$) in the final state at FCC-hh. This paper is organized as follows; the EFT effective Lagrange are detailed in the next section. The analysis steps including event generation, detector effects and event selection as well as statistical method used to obtain the limits on the coupling of anomalous CP-even and CP-odd operators are given in section III. Our results presented and discussed considering various integrated luminosity and  systematic uncertainty also in these section. Finally, conclusion is drawn in the last section.
\section{Effective CP-even and CP-odd Operators}
The elementary particles and their interactions based on the $SU(3)_c\times SU(2)_L\times U(1)_Y$ gauge symmetry are described in the Standard Model of particle physics which is a quantum field theory. All operators in the Lagrangian of the SM are restricted to the mass dimension of four or less which is consistent with Lorentz symmetry and gauge invariance. The new interactions are described in the effective-Lagrangian language as higher dimensional operators which are the residual effects on the interactions between the light degrees of freedom of the theory  after integrating out the heavy degrees of freedom.

We consider SM EFT operators as the strongly interacting light Higgs Lagrangian (SILH)  including dimension-6 operators in bar convention among the different operator bases in the literature \cite{Englert:2015hrx, Contino:2013kra, Alloul:2013naa}. In bar convention the coefficients are defined as
$\bar c \equiv c(M^2/\Lambda^2)$
where $M\equiv v, m_W$ depending on the operator normalization, and $c\sim g^2_{NP}$
is a coefficient proportional to a new physics coupling $g_{NP}$ defined at the scale $M$. Assuming the baryon and lepton number conservation, the most general form of the SILH effective Lagrangian including Higgs boson couplings that keep SM gauge symmetry is given as follows;
\begin{eqnarray}
\mathcal{L}_{eff} = \mathcal{L}_{\rm SM} + \sum_{i}\bar c_{i}O_{i}+\sum_{i} \tilde c_{i}O_{i}
\end{eqnarray}
where  $\bar c_{i}$ and $\tilde c_{i}$ are normalized Wilson coefficients of the CP-conserving and CP-violating interactions, respectively. In this study, we use the Lagrangians which describe the CP conserving and CP violating interactions between the Higgs boson and the electroweak gauge bosons as described in Ref. \cite{Alloul:2013naa}.

A part of CP-conserving operators involving the Higgs doublet  $\Phi$ of the effective Lagrangian is 
\begin{eqnarray}\label{CPC}
	\begin{split}
		\mathcal{L}_{\rm CPC} = & \
		 \frac{\bar c_{H}}{2 v^2} \partial^\mu\big[\Phi^\dag \Phi\big] \partial_\mu \big[ \Phi^\dagger \Phi \big]
		+ \frac{\bar c_{T}}{2 v^2} \big[ \Phi^\dag {\overleftrightarrow{D}}^\mu \Phi \big] \big[ \Phi^\dag {\overleftrightarrow{D}}_\mu \Phi \big] - \frac{\bar c_{6} \lambda}{v^2} \big[\Phi^\dag \Phi \big]^3
		\\
		& \
		  - \bigg[\frac{\bar c_{u}}{v^2} y_u \Phi^\dag \Phi\ \Phi^\dag\cdot{\bar Q}_L u_R
		  + \frac{\bar c_{d}}{v^2} y_d \Phi^\dag \Phi\ \Phi {\bar Q}_L d_R
		+\frac{\bar c_{l}}{v^2} y_l \Phi^\dag \Phi\ \Phi {\bar L}_L e_R
		 + {\rm h.c.} \bigg]
		\\
		&\
		 + \frac{i g\ \bar  c_{W}}{m_{W}^2} \big[ \Phi^\dag T_{2k} \overleftrightarrow{D}^\mu \Phi \big]  D^\nu  W_{\mu \nu}^k + \frac{i g'\ \bar c_{B}}{2 m_{W}^2} \big[\Phi^\dag \overleftrightarrow{D}^\mu \Phi \big] \partial^\nu  B_{\mu \nu} \\
		&\   
		+ \frac{2 i g\ \bar c_{HW}}{m_{W}^2} \big[D^\mu \Phi^\dag T_{2k} D^\nu \Phi\big] W_{\mu \nu}^k  
		+ \frac{i g'\ \bar c_{HB}}{m_{W}^2}  \big[D^\mu \Phi^\dag D^\nu \Phi\big] B_{\mu \nu}   \\
		&\
		 +\frac{g'^2\ \bar c_{\gamma}}{m_{W}^2} \Phi^\dag \Phi B_{\mu\nu} B^{\mu\nu}  
		+\frac{g_s^2\ \bar c_{g}}{m_{W}^2} \Phi^\dag \Phi G_{\mu\nu}^a G_a^{\mu\nu} 
	\end{split}
\end{eqnarray}
where $\Phi$ is Higgs sector contains a single $SU(2)_L$ doublet of fields; $\lambda$ is the Higgs quartic coupling; $g'$, $g$ and $g_s$  are coupling constant of  $U(1)_Y$, $SU(2)_L$ and $SU(3)_C$ gauge fields, respectively;  $y_u$, $y_d$ and $y_l$ are the $3\times3$ Yukawa coupling matrices in flavor space; the generators of $SU(2)_L$ in the fundamental representation are given by $T_{2k}=\sigma_k/2$ (here $\sigma_k$ are the Pauli matrices); $\overleftrightarrow{D}_\mu$ is the Hermitian derivative operators; $B^{\mu\nu}$, $W^{\mu \nu}$ and $G^{\mu\nu}$ are the electroweak and the strong field strength tensors, respectively.  

 The extra $CP$-violating operators part of the effective Lagrangian in SILH basis can be defined as,
\begin{eqnarray}
\label{CPV}
  {\cal L}_{CPV} = &\
    \frac{i g\ \tilde c_{ HW}}{\mW^2}  D^\mu \Phi^\dag T_{2k} D^\nu \Phi {\widetilde W}_{\mu \nu}^k
  + \frac{i g'\ \tilde c_{ HB}}{\mW^2} D^\mu \Phi^\dag D^\nu \Phi {\widetilde B}_{\mu \nu}
  + \frac{g'^2\  \tilde c_{ \gamma}}{\mW^2} \Phi^\dag \Phi B_{\mu\nu} {\widetilde B}^{\mu\nu}\\
 &\
  +\!  \frac{g_s^2\ \tilde c_{ g}}{\mW^2}      \Phi^\dag \Phi G_{\mu\nu}^a {\widetilde G}^{\mu\nu}_a
  \!+\!  \frac{g^3\ \tilde c_{3W}}{\mW^2} \epsilon_{ijk} W_{\mu\nu}^i W^\nu{}^j_\rho {\widetilde W}^{\rho\mu k}
  \!+\!  \frac{g_s^3\ \tilde c_{ 3G}}{\mW^2} f_{abc} G_{\mu\nu}^a G^\nu{}^b_\rho {\widetilde G}^{\rho\mu c} \ \nonumber
\end{eqnarray}
where \begin{eqnarray}
  \widetilde B_{\mu\nu} = \frac12 \epsilon_{\mu\nu\rho\sigma} B^{\rho\sigma} \ , \quad
  \widetilde W_{\mu\nu}^k = \frac12 \epsilon_{\mu\nu\rho\sigma} W^{\rho\sigma k} \ , \quad
  \widetilde G_{\mu\nu}^a = \frac12 \epsilon_{\mu\nu\rho\sigma} G^{\rho\sigma a} \ \nonumber 
\end{eqnarray}
are the dual field strength tensors.

The SILH bases of CP-conserving and CP-violating dimension-6 operators given in Eq.\ref{CPC} and Eq.\ref{CPV} can be defined in terms of the mass eigenstates after electroweak symmetry breaking. In the mass basis and in the unitarity gauge, the general effective Lagrangian associated with to the 3-point interactions involving at least one Higgs boson as is follows \begin{eqnarray}\label{massb}
 {\cal L} &= &
    - \frac{m_{\sss H}^2}{2 v} g^{(1)}_{\sss hhh}h^3 + \frac{1}{2} g^{(2)}_{\sss hhh} h\partial_\mu h \partial^\mu h\nonumber
     - \frac{1}{4} g_{\sss hgg} G^a_{\mu\nu} G_a^{\mu\nu} h
    - \frac{1}{4} \tilde g_{\sss hgg} G^a_{\mu\nu} \tilde G^{\mu\nu} h
    - \frac{1}{4} g_{\sss h\gamma\gamma} F_{\mu\nu} F^{\mu\nu} h    \\ \nonumber 
    &-& \frac{1}{4} \tilde g_{\sss h\gamma\gamma} F_{\mu\nu} \tilde F^{\mu\nu} h
    - \frac{1}{4} g_{\sss hzz}^{(1)} Z_{\mu\nu} Z^{\mu\nu} h
    - g_{\sss hzz}^{(2)} Z_\nu \partial_\mu Z^{\mu\nu} h
    + \frac{1}{2} g_{\sss hzz}^{(3)} Z_\mu Z^\mu h
    - \frac{1}{4} \tilde g_{\sss hzz} Z_{\mu\nu} \tilde Z^{\mu\nu} h\\ \nonumber 
    &-& \frac{1}{2} g_{\sss haz}^{(1)} Z_{\mu\nu} F^{\mu\nu} h
    - \frac{1}{2} \tilde g_{\sss haz} Z_{\mu\nu} \tilde F^{\mu\nu} h
    - g_{\sss haz}^{(2)} Z_\nu \partial_\mu F^{\mu\nu} h
    - \frac{1}{2} g_{\sss hww}^{(1)} W^{\mu\nu} W^\dag_{\mu\nu} h\\ \nonumber
   & - &\Big[g_{\sss hww}^{(2)} W^\nu \partial^\mu W^\dag_{\mu\nu} h + {\rm h.c.} \Big]
    +  g (1-\frac12 \bar c_{\sss H}) \mW W_\mu^\dag  W^\mu h 
    - \frac{1}{2} \tilde g_{\sss hww} W^{\mu\nu} \tilde W^\dag_{\mu\nu} h\\
    &-& \bigg[ 
      \tilde y_u \frac{1}{\sqrt{2}} \big[{\bar u} P_R u\big] h +
      \tilde y_d \frac{1}{\sqrt{2}} \big[{\bar d} P_R d\big] h +
      \tilde y_\ell \frac{1}{\sqrt{2}} \big[{\bar \ell} P_R \ell\big] h
     + {\rm h.c.} \bigg] \ ,
\end{eqnarray}
where $G_{\mu\nu}$, $Z_{\mu\nu}$ and $F_{\mu\nu}$ are the field strength tensors of gluon, $Z$-boson and photon, respectively; $m_H$ represent the mass of the Higgs boson; the effective couplings in gauge basis defined as dimension-6 operators are given in Table I in which $a_H$ ($g_H$) coupling is the SM contribution to the Higgs boson to two photons (gluons) vertex at loop level.

\begin{table}[h]
\caption{The relations between Lagrangian  parameters in the mass basis (Eq.\ref{massb}) and the Lagrangian in gauge  basis (Eqs. \ref{CPC} and \ref{CPV}). ($c_W\equiv\cos \theta_W$, $s_W\equiv\sin \theta_W$)}  
\begin{ruledtabular}\label{mtable}
\begin{tabular}{ll}
      $g_{ hhh}^{(1)}$ = $1 + \frac78 \bar c_{\sss 6} - \frac12 \bar c_{\sss H}$ & $\tilde g_{hgg}$ =$ - \frac{4 \tilde c_{ g} g_s^2 v}{\mW^2}$\\
  $g_{\sss hhh}^{(2)}$ =$\frac{g}{\mW} \bar c_{\sss H}$& $\tilde g_{ h\gamma\gamma}$=$ -\frac{8 g \tilde c_{ \gamma} \sW^2}{\mW}$ \\
$g_{hgg}$ = $g_{H} - \frac{4 \bar c_{ g} g_s^2 v}{\mW^2}$                               & $\tilde g_{ hzz}$ =$\frac{2 g}{\cW^2 \mW} \Big[ \tilde c_{ HB} \sW^2 - 4 \tilde c_{ \gamma} \sW^4 + \cW^2 \tilde c_{ HW}\Big]$  \\
$g_{h\gamma\gamma}$= $a_{ H} - \frac{8 g \bar c_{ \gamma} \sW^2}{\mW}$ &$\tilde g_{ h\gamma z}$ =$\frac{g \sW}{\cW \mW} \Big[  \tilde c_{HW} - \tilde c_{ HB} + 8 \tilde c_{ \gamma} \sW^2\Big]$   \\
$g^{(1)}_{ hzz}$= $\frac{2 g}{\cW^2 \mW} \Big[ \bar c_{HB} \sW^2 - 4 \bar c_{ \gamma} \sW^4 + \cW^2 \bar c_{ HW}\Big]$& $\tilde g_{ hww}$= $\frac{2 g}{\mW} \tilde c_{ HW}$                               \\
$g^{(2)}_{ hzz}$= $\frac{g}{\cW^2 \mW} \Big[(\bar c_{ HW} +\bar c_{ W}) \cW^2  + (\bar c_{ B} + \bar c_{ HB}) \sW^2 \Big]$&   $g^{(3)}_{hzz}$=  $\frac{g \mW}{\cW^2} \Big[ 1 -\frac12 \bar c_{H} - 2 \bar c_{T} +8 \bar c_{\gamma} \frac{\sW^4}{\cW^2} \Big]$  \\
     $g^{(1)}_{ h\gamma z}$= $\frac{g \sW}{\cW \mW} \Big[  \bar c_{ HW} - \bar c_{HB} + 8 \bar c_{ \gamma} \sW^2\Big]$& $g^{(2)}_{h\gamma z}$= $\frac{g \sW}{\cW \mW} \Big[  \bar c_{HW} - \bar c_{ HB} - \bar c_{ B} + \bar c_{ W}\Big]$  \\
    $g^{(1)}_{hww}$=$\frac{2 g}{\mW} \bar c_{HW}$&$g^{(2)}_{ hww}$=$\frac{g}{\mW} \Big[ \bar c_{ W} + \bar c_{ HW} \Big]$\\
     \end{tabular}
\end{ruledtabular}
\end{table}

The $pp\to\gamma\gamma$+n-jet process is sensitive to interactions between the Higgs boson and two photons and between the Higgs boson and two gluons ($g_{h\gamma\gamma}$ and $g_{hgg}$) and the couplings of a quark pair to single Higgs field ($\tilde y_u$ and $\tilde y_d$) in the mass basis. This process is also sensitive to the four Wilson coefficients in the gauge basis:  $\bar c_{\gamma}$, $\bar{c}_{g}$, $\tilde{c}_{\gamma}$ and $\tilde{c}_{g}$ related to Higgs-gauge boson couplings and also effective fermionic couplings in the gauge basis. Due to the small Yukawa couplings of the first and second generation fermions, we neglect the effective fermionic couplings. 
\begin{table}
\caption{The  list of subproceses which contribute to $pp\to\gamma\gamma$+n-jet (n-jet=0,1 and 2) processes at LO (where $q=u,c$ and $q'= d ,s, b$). }  
\begin{ruledtabular}\label{sprocess}
\begin{tabular}{lll}
Process&Subprocesses& Number of subprocesses\\\hline
&$gg\to \gamma\gamma$&1\\
$pp\to\gamma\gamma$&$q'\bar{q'}\to \gamma\gamma$&6\\
&$q\bar q\to \gamma\gamma$&4\\\hline
&$g g \to \gamma\gamma g$&1\\
&$q'\bar {q'}\to \gamma\gamma g$&6\\
&$g q\to \gamma\gamma q $&4\\
$pp\to\gamma\gamma$+1 jet&$g \bar{q} \to \gamma\gamma \bar{q}$&4\\
&$g q'\to \gamma\gamma q' $&4\\
&$g \bar{q'} \to \gamma\gamma \bar{q'}$&4\\
&$q\bar q\to \gamma\gamma g$&4\\\hline
&$gg\to \gamma\gamma g g$&1\\
&$gg\to \gamma\gamma q \bar q $&2\\
&$gg\to \gamma\gamma q' \bar {q'} $&2\\
&$qq\to \gamma\gamma q q$&4\\
&$\bar q\bar q\to \gamma\gamma \bar q \bar q$&4\\
&$ q\bar q\to \gamma\gamma  q \bar q$&12\\
&$ q\bar q\to \gamma\gamma  g g$&4\\
&$ q\bar q\to \gamma\gamma  q' \bar {q'}$&32\\
&$q'q'\to \gamma\gamma q' q'$&4\\
&$\bar{q'}\bar {q'}\to \gamma\gamma \bar {q'} \bar {q'}$&4\\
$pp\to\gamma\gamma$+2 jets&$q'\bar {q'}\to \gamma\gamma q \bar {q}$&72\\
&$q'\bar {q'}\to \gamma\gamma q' \bar {q'}$&16\\
&$q'\bar {q'}\to \gamma\gamma g g $&6\\
&$ q q'\to \gamma\gamma  q q'$&48\\
&$ q\bar {q'}\to \gamma\gamma  q \bar {q'}$&48\\
&$ q'\bar q\to \gamma\gamma  q' \bar q$&48\\
&$ \bar q \bar {q'}\to \gamma\gamma  \bar q \bar{ q'}$&48\\
&$ g q\to \gamma\gamma  g q$&4\\
&$ g q'\to \gamma\gamma  g q'$&4\\
&$ g\bar {q}\to \gamma\gamma  g \bar {q}$&4\\
&$ g\bar {q'}\to \gamma\gamma  g \bar {q'}$&4\\
 \end{tabular}
\end{ruledtabular}
\end{table}
 \section{Signal and Background Analysis}
As is well known, the analysis of di-photon production in hadron-hadron collision is very attractive not only because it is the main background source of the Higgs boson in the di-photon decay channel but also tool to look for physics beyond the SM. Therefore, we focus on $pp\to\gamma\gamma$+n-jet (where n-jet=0,1 and 2) production mechanism using Monte Carlo simulations with leading order (LO) in \verb|MadGraph5_aMC@NLO v2.6.7| \cite{Alwall:2014hca} to determine the sensitivity interval for dimension-6 operators. In accordance with this purpose, the effective Lagrangian of the SM EFT in Eq.(\ref{massb}) is implemented into the \verb|MadGraph5_aMC@NLO| using FeynRules \cite{Alloul:2013bka} and UFO \cite{Degrande:2011ua} framework. In Table \ref{sprocess}, we give the detailed list of 409 subprocesses which contribute to di-doson processes with zero, one and two external jets at LO considered in this study. Signal events are generated for $pp\to\gamma\gamma$+n-jet (where n-jet=0,1 and 2) process with different values of a given Wilson coefficients $\bar{c}_{g}$, $\tilde{c}_{g}$, $\bar{c}_{\gamma}$ and $\tilde{c}_{\gamma}$  includes SM contribution as well as interference between effective couplings and SM contributions ($S+B_{1}$).
As relevant backgrounds, we consider not only the SM contribution ($B_{1}$) with the same final state of the signal process, but also  two photon in association with a top–antitop pair $B_2$ ($pp\to t\bar t\gamma\gamma$) and vector bosons $B_3$($pp\to W\gamma\gamma$), $B_4$($pp\to Z\gamma\gamma$) production processes. The signal ($S+B_{1}$) and SM ($B_{1}$) background events at parton level with up to two additional partons are generated in the final state in \verb|MadGraph5_aMC@NLO v2.6.7|. The zero, one and two parton events are merged using the MLM matching scheme \cite{Mangano:2006rw}. All signal and relevant background events passed through the Pythia8 \cite{Sjostrand:2006za} for parton showering and hadronization. Jets are clustered by using FastJet \cite{Cacciari:2011ma} with anti-$k_t$ algorithm where a cone radius is set as $\Delta R$ = 0.4 \cite{Cacciari:2008gp}. The detector responses are taken into account with FCC-hh detector card in \verb|Delphes 3.4.2| \cite{deFavereau:2013fsa} package. It is known that the pileup effects would be a serious problem at a high energy and high luminosity runs at hadron colliders. However, pile-up effects are not taken into account in our study. Finally, all events are analysed by using the ExRootAnalysis utility \cite{exroot} with ROOT 6.16 \cite{Brun:1997pa}.

One can get preliminary information to see the effect of cross section for the $pp\to\gamma\gamma$+n-jet signal process at generator level as function of  $\bar{c}_{g}$, $\tilde{c}_{g}$, $\bar{c}_{\gamma}$ and $\tilde{c}_{\gamma}$ couplings. The  total cross section for the $pp\to\gamma\gamma$+n-jet signal processes as a function of CP-conserving ($\bar c_{i}$), CP-violating ($\tilde c_{i}$) couplings and SM parts can be expected to be written as follows; 
\[\sigma_{tot} (\bar {c_{i}},\tilde {c_{i}}) = \bar c_{i}^2\sigma_{cpc}+\tilde {c_{i}}^2\sigma_{cpv}+\bar c_{i}\sigma_{int}+\sigma_{SM}\]
where $\sigma_{SM}$ is SM background cross section which is the same final state of the signal process; $\sigma_{cpc}$ and $\sigma_{cpv}$ are CP-conserving and CP-violating dimension-6 operators contributions to the total cross section, respectively  while $\sigma_{int}$ is the interference contribution between CP-conserving operators with the dimension-4 SM operators. Notice that due to nature of CP-violating dimension-6 operators there is no interference between those and the SM contributing to the total cross section. Therefore, we generate 36 samples when studying two Wilson coefficients simultaneously to parametrize the cross section function. Parametrized the total cross section in $pb$ as a function of $\bar{c}_{\gamma}$, $\tilde{c}_{\gamma}$  and $\bar{c}_{g}$, $\tilde{c}_{g}$ are 
\[\sigma_{tot} (\bar c_{\gamma},\tilde {c_{\gamma}}) = 3.64\times10^5 \bar {c_{\gamma}}^2+8.49x10^5\tilde {c_{\gamma}}^2+2.03\times10^4\bar c_{\gamma}+4.91\times10^3\]
\[\sigma_{tot} (\bar c_{g},\tilde {c_{g}}) = 2.91\times10^7 \bar {c_{g}}^2+7.75x10^7\tilde {c_{g}}^2-1.60\times10^4\bar c_{g}+4.91\times10^3\]

\begin{figure}[hbt!]
\includegraphics[scale=0.35]{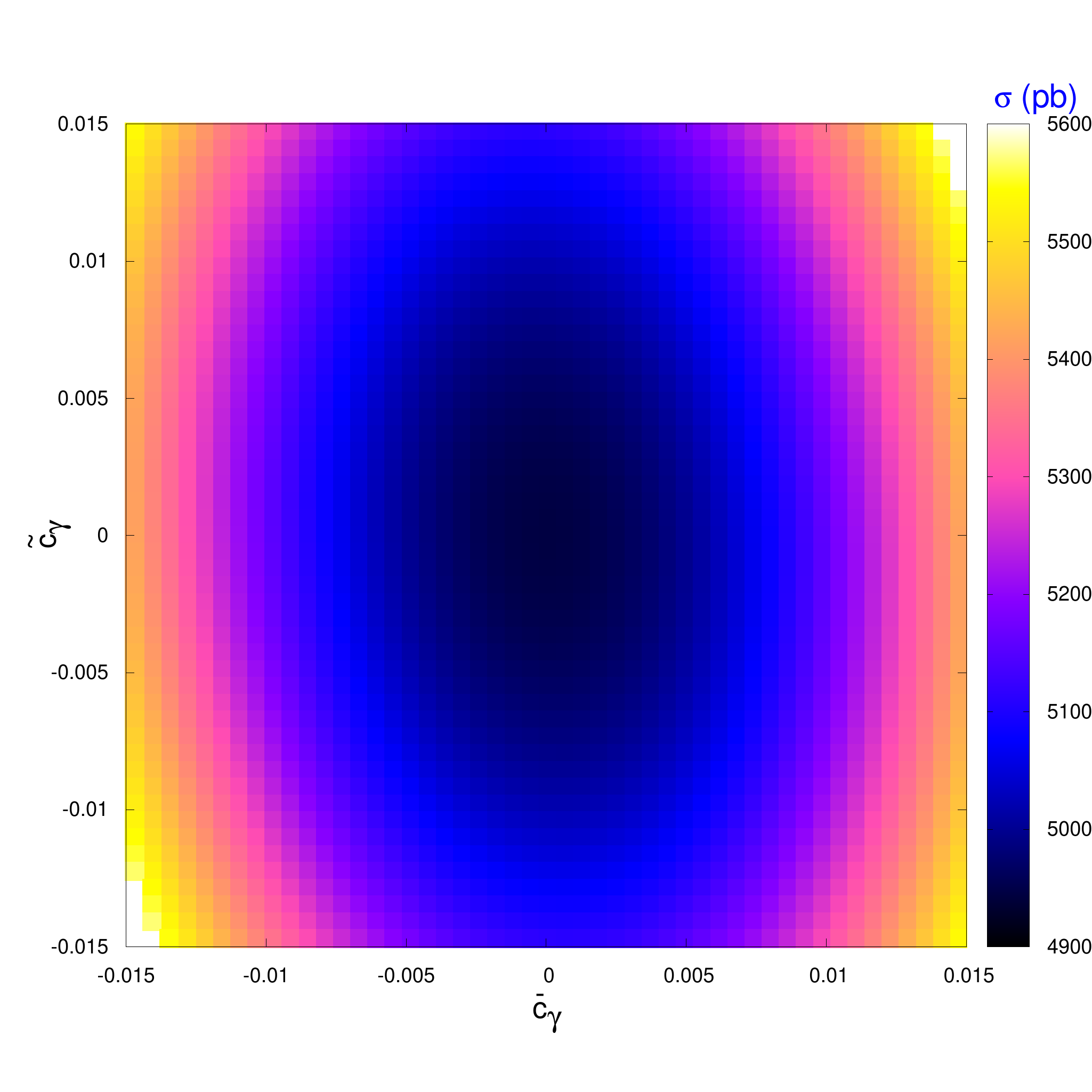} \includegraphics[scale=0.35]{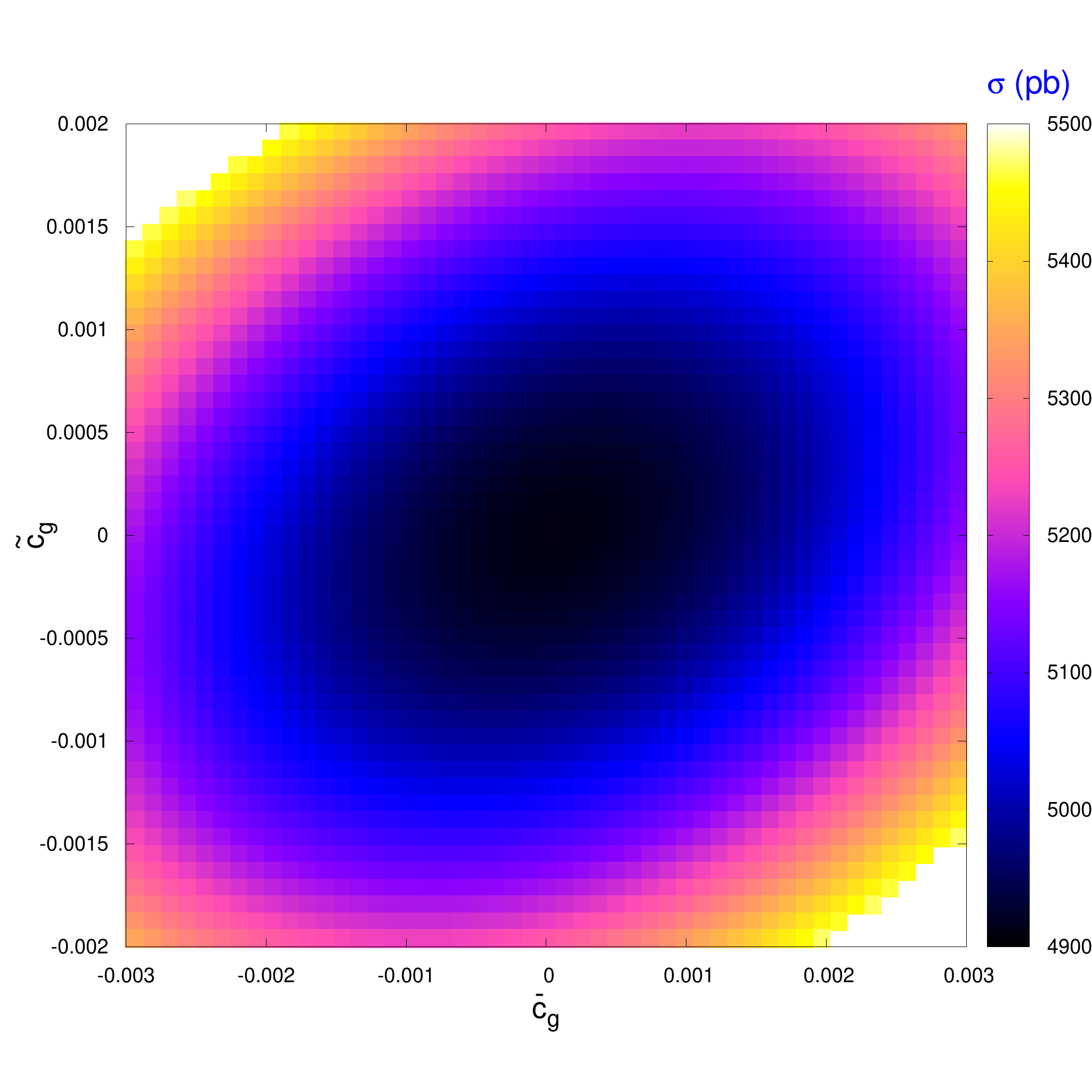} 
\caption{ Variation of the total cross section as a function of  ${c}_{\gamma}$ and $\tilde{c}_{\gamma}$ couplings (on the left) and ${c}_{g}$ and $\tilde{c}_{g}$ (on the right) for the $pp\to\gamma\gamma$+n-jet (where n-jet=0, 1 and 2) process at the FCC-hh with $\sqrt s$=100~TeV. \label{fig1}}
\end{figure}

Finally, the method is validated by comparing the cross sections obtained with the parametrization function to the obtained cross section with event samples generated at the specific point in parameter space. 
Fig.\ref{fig1} shows  the variation of cross sections of $pp\to\gamma\gamma$+n-jet process in $\bar{c}_{\gamma}$-$\tilde{c}_{\gamma}$ couplings plane on the left panel and $\bar{c}_{g}$-$\tilde{c}_{g}$  couplings plane on the right panel at FCC-hh with 100 TeV center of mass energy. The photon and jet transverse momentum grater than 15 GeV at the generator level is required to calculate these cross sections. In these figures, the effective couplings under consideration are varied (a two-dimensional scan), while the others are fixed to zero. These figures also lead us the parameter range we study for detailed analysis.
\begin{figure}[hbt!]
\includegraphics[scale=0.8]{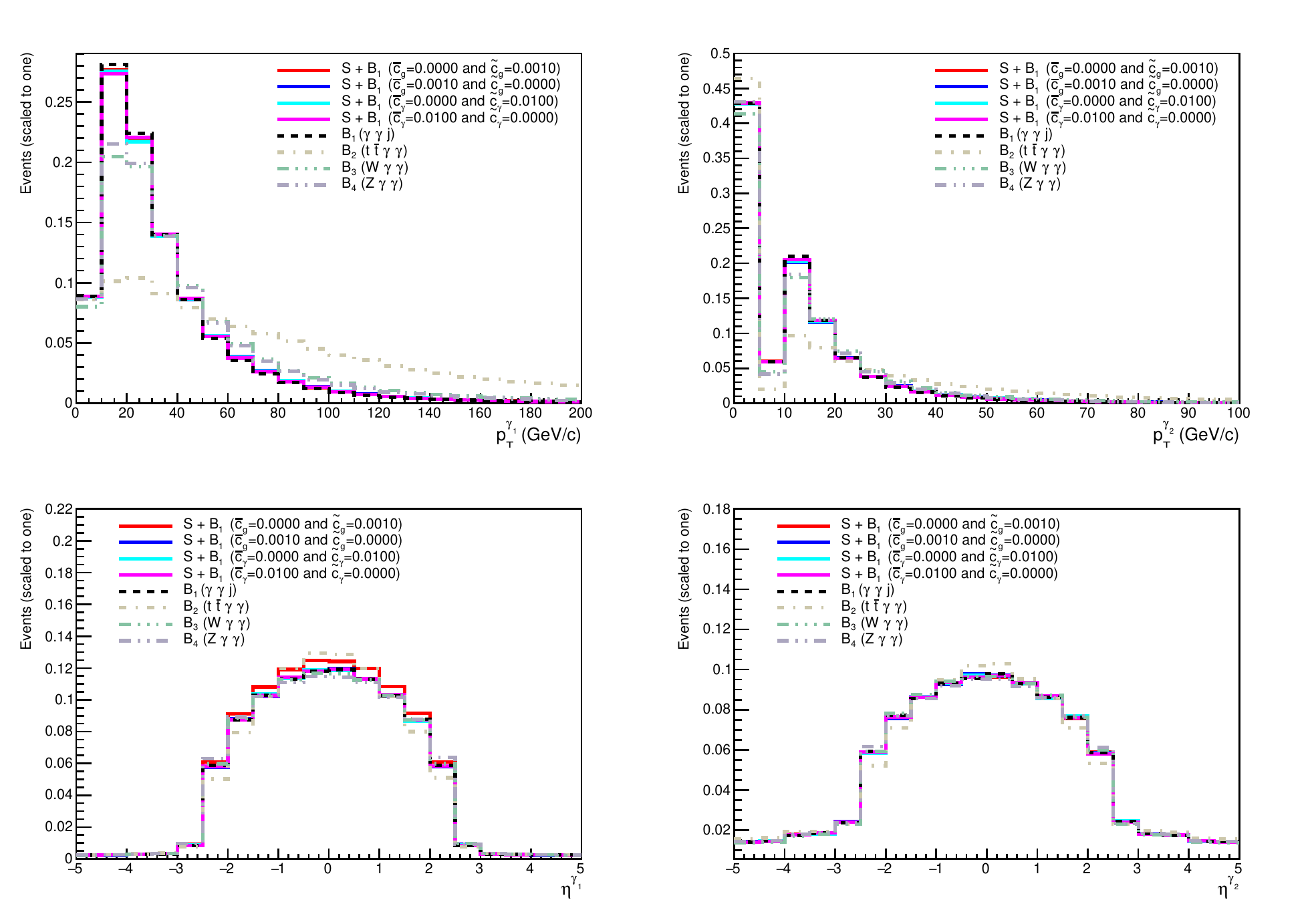} 
\caption{ The transverse momentum (in the first row) and pseudo-rapidity distribution (in the second row) of the leading and sub-leading (left-to-right) photons for the $pp\to\gamma\gamma$+n-jet (where n-jet=0, 1 and 2) signal process and their relevant backgrounds at the FCC-hh with $\sqrt s$=100~TeV. These distributions are normalized to one.\label{fig2}}
\end{figure} 

\begin{figure}[hbt!]
\includegraphics[scale=0.85]{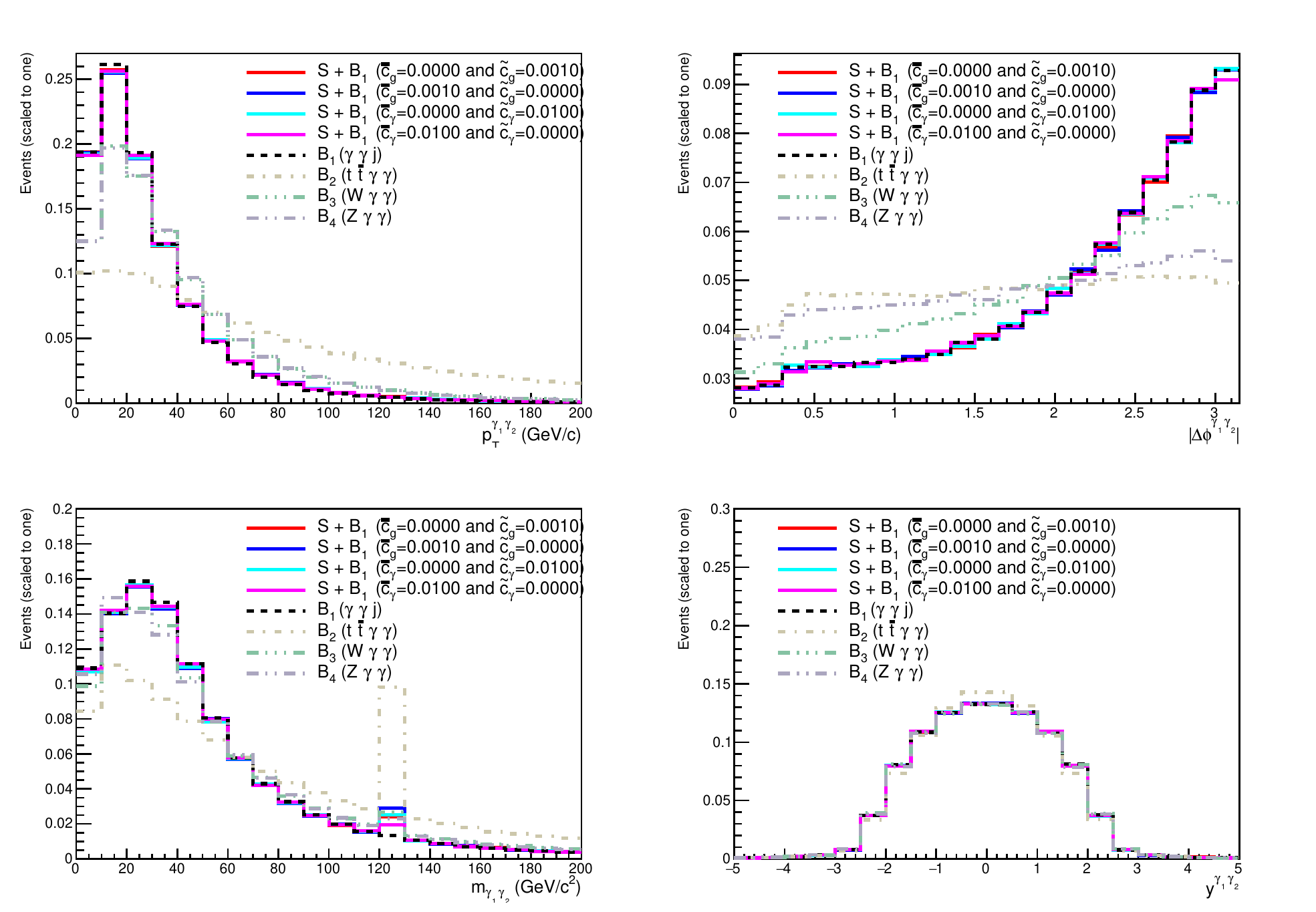} 
\caption{ The transverse momentum and the azimuthal angle between the leading and sub-leading  photon (in the first row), invariant mass and rapidity distribution (in the second row) (left-to-right ) of the diphoton system for $pp\to\gamma\gamma$+n-jet (where n-jet=0, 1 and 2) signal process and their relevant backgrounds at the FCC-hh with $\sqrt s$=100~TeV. These distributions are normalized to one.\label{fig3}}
\end{figure} 

\begin{figure}[hbt!]
\includegraphics[scale=0.40]{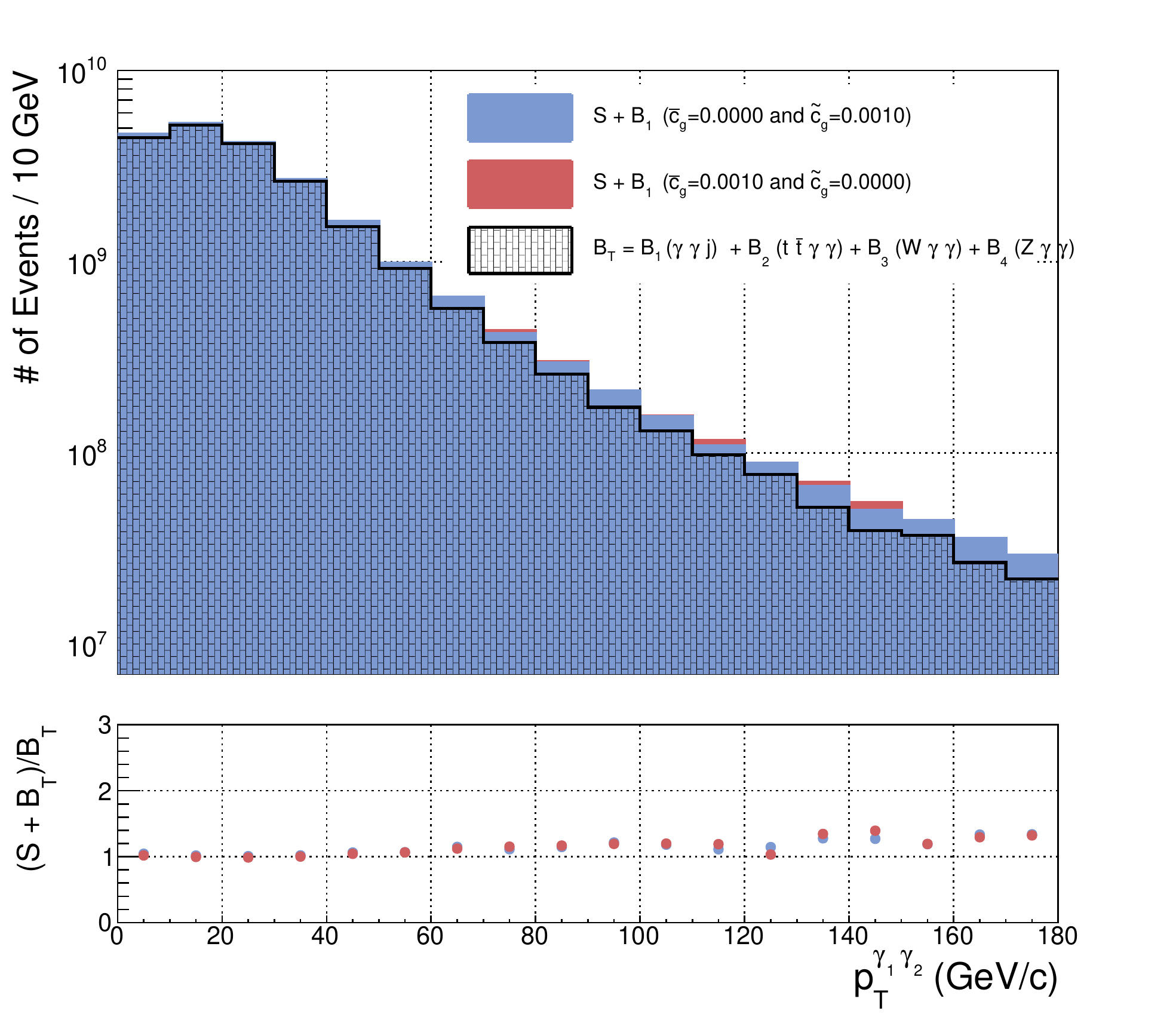} \includegraphics[scale=0.40]{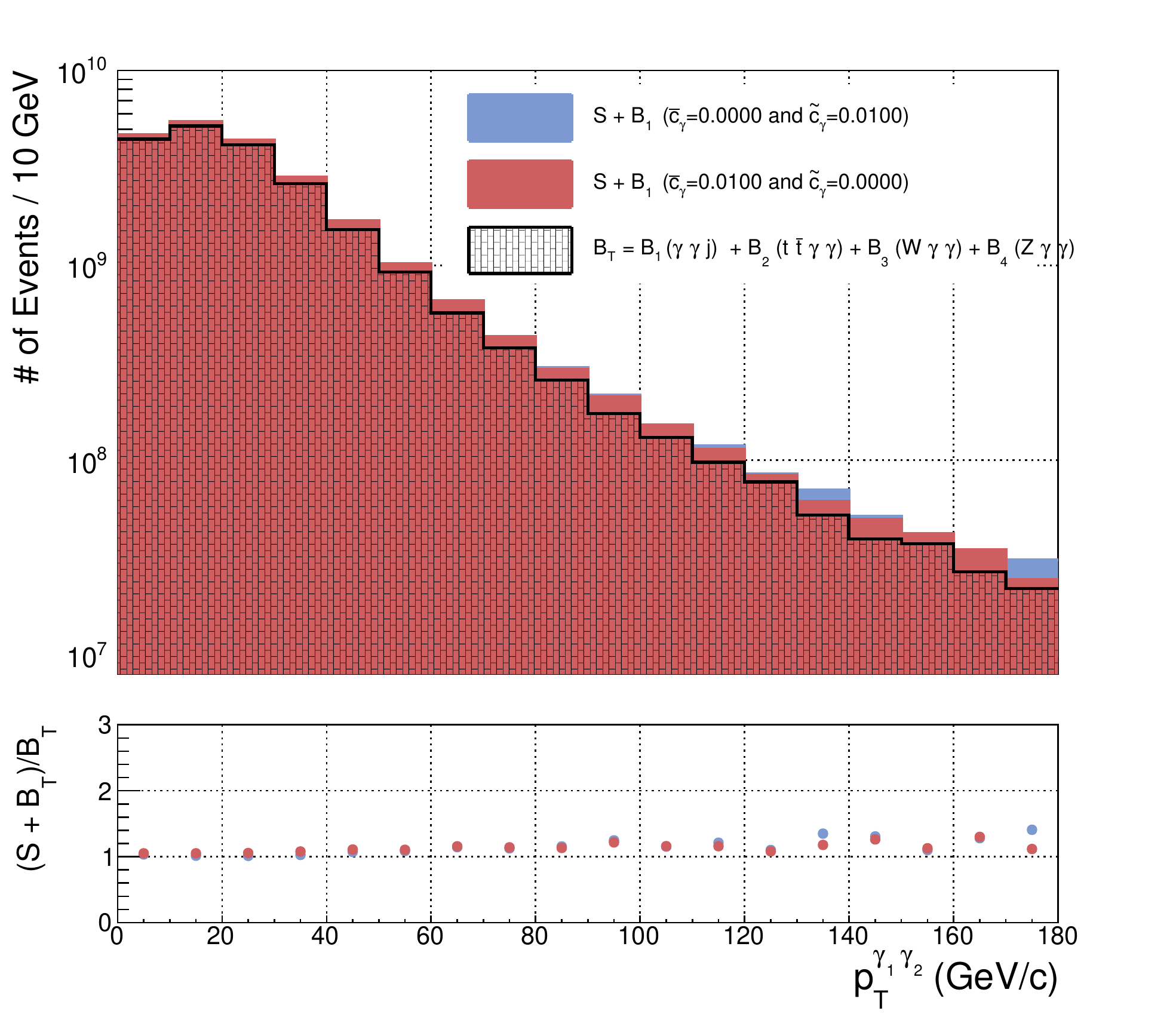} 
\caption{ After pre-selection and Kinematic cuts, the number of events as function of the transverse momentum for the diphoton system for $pp\to\gamma\gamma$+n-jet (where n-jet=0, 1 and 2) signal process and their relevant backgrounds at the FCC-hh with $\sqrt s$=100~TeV and $L_{int}=1$ ab$^{-1}$.\label{fig4}}
\end{figure} 

Since unique signature of our signal process requires at least two photons, events with $N_{\gamma}>1$ as well as their transverse momenta grater than 0.5 GeV is defined to be pre-selection for the detailed analysis. The photons are ordered according to their transverse momentum, i.e., $p_T^{\gamma_1} > p_T^{\gamma_2}$. 
The first row of Fig.\ref{fig2} shows transverse momentum ($p_T^{\gamma}$) for  the leading and sub-leading photon for four different set in which only one Wilson coefficient is non-zero for signal and relevant SM backgrounds while second row shows pseudo-rapidity distributions.
Based on these distributions, deviations from SM backgrounds start to be visible around $p_T^{\gamma_1}>$ 40 GeV, $p_T^{\gamma_2}>$ 30 GeV  and $|\eta^{\gamma_{1,2}}| < 2.5$. Instead of using fix cut in $p_T^{\gamma_1,\gamma_2}$ which result in distortion at the low end of the invariant mass spectrum of two photon, we use a cut on variables $p_T^{{\gamma_1}({\gamma_2})}/m_{\gamma_1\gamma_2}$ to be grater than 1/3 (1/4). We also reconstruct  the candidate Higgs from two photons and plot its transverse momentum ($p_T^{\gamma_1\gamma_2}$), the azimuthal angle difference between two photon $\Delta \phi ^{\gamma_1\gamma_2}=\phi_{\gamma_1}-\phi_{\gamma_2}$, its invariant mass $m_{\gamma_1\gamma_2}$ and rapidity $y^{\gamma_1\gamma_2}=1/2 \text{ln} [(E+p_z)/(E-p_z)]$, (where $E$ is the energy and $p_z$ is the $z$-component of the momentum) as seen in Fig.\ref{fig3} for signal and relevant backgrounds. Among these variables, both invariant mass and transverse momentum of diphoton system are more sensitive to obtain constrain on the Wilson coefficients of dimension-six CP-even or CP-odd operators of Higgs boson to photons and gluon vertices. The minimum distance between each photon is also required to satisfy $\Delta R(\gamma_i,\gamma_j)= \left[(\Delta\phi_{\gamma_i,\gamma_j}])^2+(\Delta\eta_{\gamma_i,\gamma_j}])^2\right]^{1/2} > 0.4$ where $\Delta\phi_{\gamma_i,\gamma_j}$  and $\Delta\eta_{\gamma_i,\gamma_j}$ are azimuthal angle and the pseudo rapidity difference between any two photons.  Fig.\ref{fig4} shows the distributions of the transverse momentum of the reconstructed Higgs boson from two leading photon after applying following cuts; $p_T^{\gamma_1}/m_{\gamma_1\gamma_2}>1/3$ , $p_T^{\gamma_2}/m_{\gamma_1\gamma_2}>1/4$ and $|\eta^{\gamma_{1}}|< 2.5,|\eta^{\gamma_{2}}|< 2.5$ and $\Delta R(\gamma_1,\gamma_2) > 0.4$. Since the invariant mass of the di-photon system, $m_{\gamma_1 \gamma_2}$, is peaked around 125 GeV in both the signals and relevant backgrounds, we select events in the range of $|m_{\gamma1 \gamma2}-125|<4$ GeV. Finally events in which transverse momentum of two-photon system greater than 60 GeV are used to obtain limits on the anomalous Higgs effective couplings. The definition of used cuts in the analysis are summarized in Table \ref{cuts}. We presented the number of events after each cuts used in the analysis for signal ($S+B_1$( $\bar{c}_{g}$=0.001), $S+B_1$( $\tilde{c}_{g}$=0.001), $S+B_1$( $\bar{c}_{\gamma}$=0.01) and  $S+B_1$( $\tilde{c}_{\gamma}$=0.01)) and relevant SM backgrounds ($B_1$,$B_2$,$B_3$ and $B_4$) in Table \ref{eff}. The number events in this table are normalized to the cross section of each process times the integrated luminosity, $L_{int}$=1  ab$^{-1}$.

\begin{table}[hbt!]
\caption{Summary of event selection and definitions of kinematical cuts used for the analysis of signal and background events. \label{cuts}}

\begin{tabular}{lccc}
 \hline\hline 
 && Cuts  \\ \hline\hline 
Pre-selection && $N_\gamma \geqslant 2$  \\
Kinematics& &$p_T^{{\gamma_1}({\gamma_2})}/m_{\gamma_1\gamma_2} >1/3 (1/4)$,\\
&&$|\eta^{\gamma_{1}}| < 2.5$, $|\eta^{\gamma_{2}}| < 2.5$  \\
&&$\Delta R(\gamma_1,\gamma_2) > 0.4$  \\
Higgs-reconstruction && 121 GeV $< m_{\gamma\gamma}< 129 $ GeV \\
& &$p_T^{{\gamma_1\gamma_2} }> 60$ GeV \\\hline \hline
\end{tabular}
\end{table}

\begin{table}[hbt!]
\caption{The number of events after applied cuts for signal and background processes. The numbers are normalized to the cross section of each process times the integrated luminosity, $L_{int}$=1  ab$^{-1}$. }  
\begin{ruledtabular}\label{eff}
\begin{tabular}{llcc}
Processes&Pre-Selection&Kinematics&Higgs-reconstruction  \\\hline
$S+B_1$( $\bar{c}_{g}$=0.001)&$4.29\times10^{9}$ & $2.10\times10^{9}$&$3.30\times10^{7}$\\
$S+B_1$( $\tilde{c}_{g}$=0.001)&$4.35\times10^{9}$ & $2.14\times10^{9}$&$2.52\times10^{7}$\\
$S+B_1$( $\bar{c}_{\gamma}$=0.01)&$4.48\times10^{9}$ & $2.20\times10^{9}$&$1.64\times10^{7}$\\
$S+B_1$( $\tilde{c}_{\gamma}$=0.01)&$4.35\times10^{9}$ & $2.14\times10^{9}$&$2.72\times10^{7}$\\
$B_1$&$4.16\times10^{9}$ & $2.05\times10^{9}$&$5.76\times10^{6}$\\
$B_2$&$3.86\times10^{5}$ & $1.56\times10^{5}$&$2.10\times10^{4}$\\
$B_3$&$7.25\times10^{5}$ & $3.47\times10^{5}$&$5.34\times10^{3}$\\
$B_4$&$7.64\times10^{5}$ & $3.53\times10^{5}$&$4.25\times10^{3}$\\
  \end{tabular}
\end{ruledtabular}
\end{table}
One can construct a $\chi^{2}$ test using the transverse momentum distributions of diboson system of the $pp\to\gamma\gamma$+n-jet signal process and relevant SM backgrounds in the range of 121 GeV $< m_{\gamma\gamma}< 129 $ GeV and find limits Wilson coefficients at 95\% C.L.. as
\begin{eqnarray}
\chi^{2} (\bar{c_i})=\sum_i^{n_{bins}}\left(\frac{N_{i}^{NP}(\bar{c_i})-N_{i}^{B}}{N_{i}^{B}\Delta_i}\right)^{2}
\end{eqnarray}
where $N_i^{NP}$ is the total number of events in the existence of effective couplings ($S$) , $N_i^B$ is the total number of relevant SM background events in $i$th bin. $\Delta_i=\sqrt{\delta_{sys}^2+\frac{1}{N_i^B}}$ is the combined systematic ($\delta_{sys}$) and statistical errors in each bin. In this analysis, we focused on the main coefficients contributing to $pp\to\gamma\gamma$+n-jet signal process i.e., $\bar{c}_{g} $, $\tilde{c}_{g} $, $\bar{c}_{\gamma} $ and $\tilde{c}_{\gamma} $ couplings.

\begin{figure}[hbt!]
\includegraphics[scale=0.40]{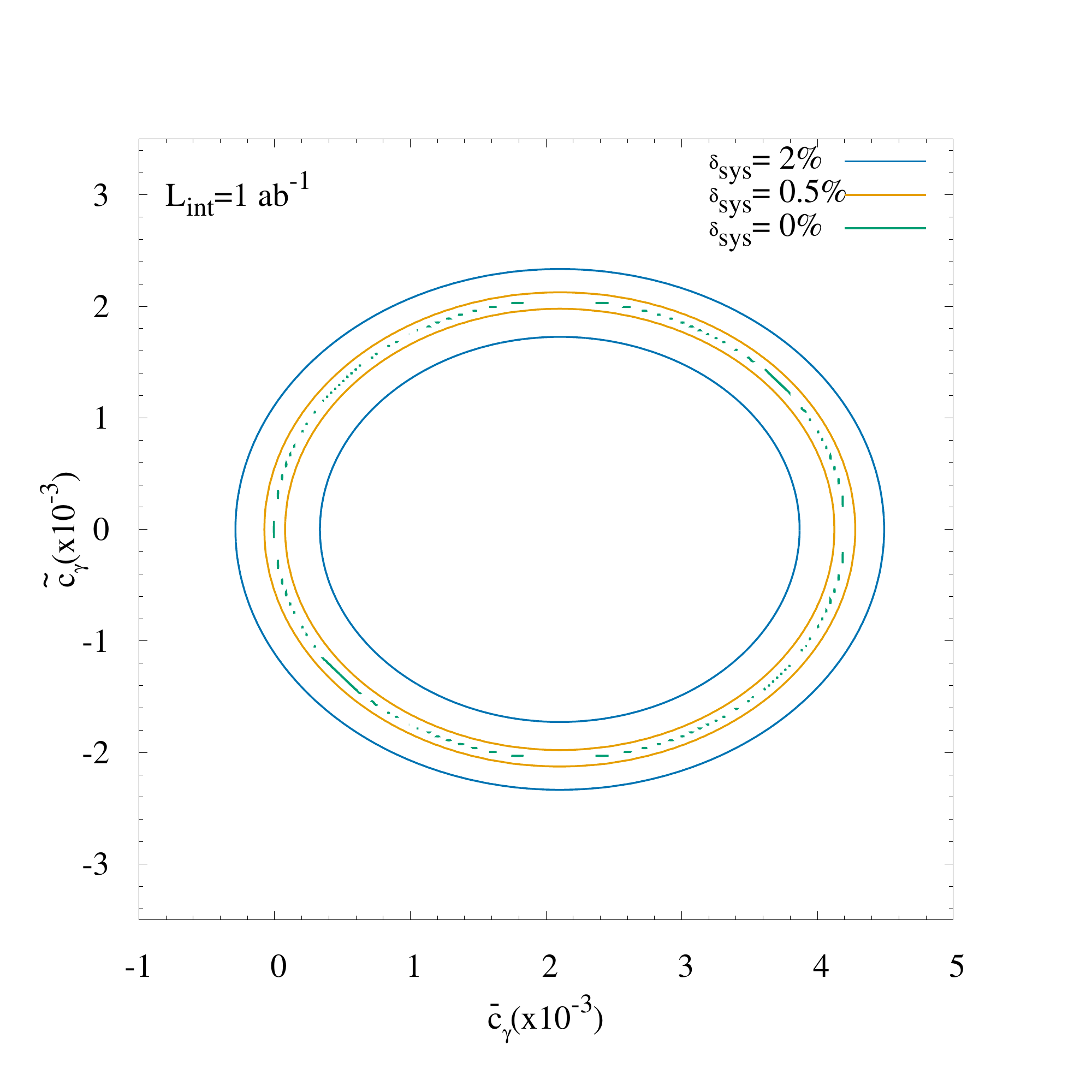} \includegraphics[scale=0.40]{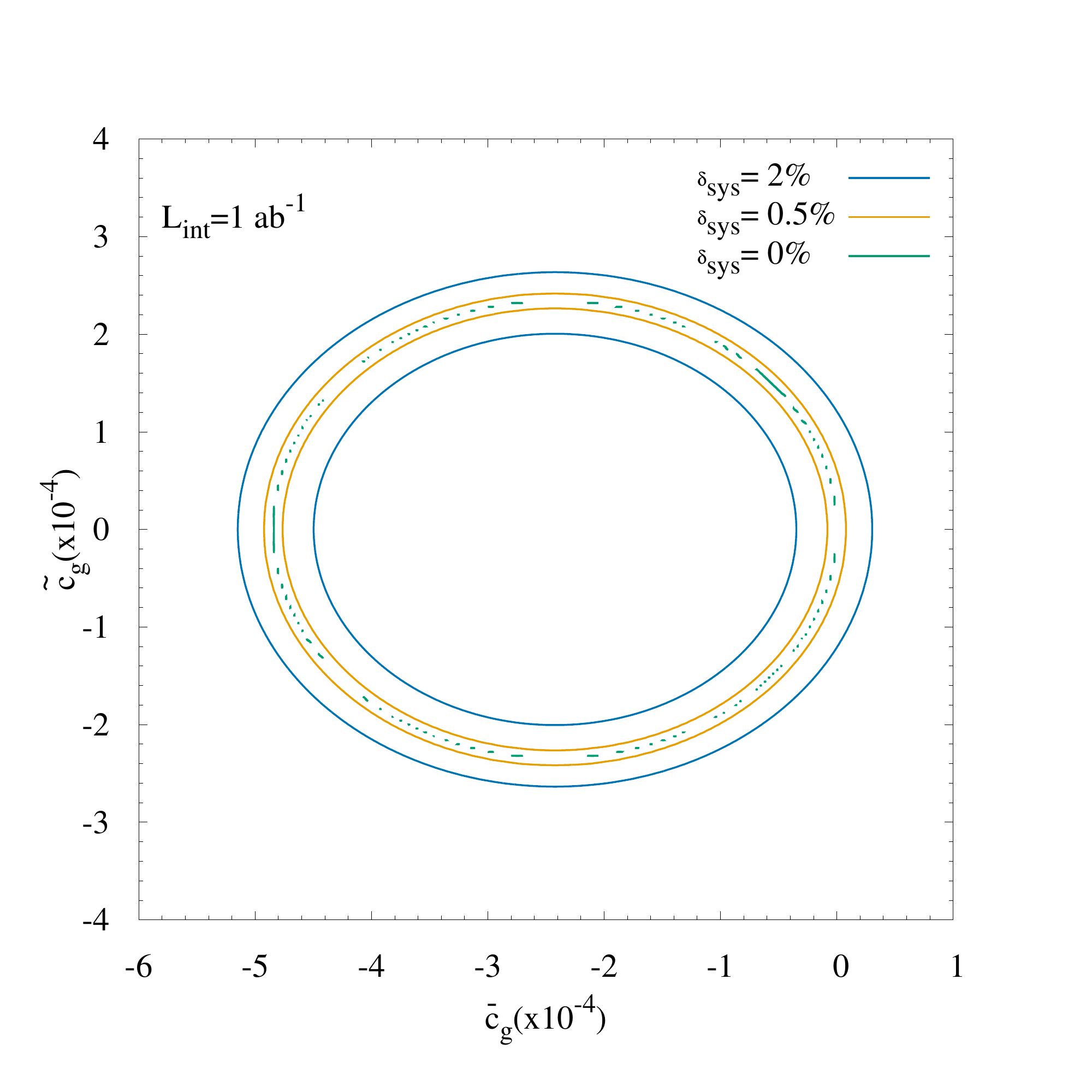} 
\caption{Two-dimensional 95\% C.L. intervals in plane for $\bar{c}_{\gamma}$-$\tilde{c}_{\gamma}$ (on the left) and $\bar{c}_{g}$-$\tilde{c}_{g}$ (on the right) with taking  $\delta_{sys}$=0, 0.5\% and 2\% of systematic errors at $L_{int}=1$  ab$^{-1}$ (on the right) for $\sqrt s$=100~TeV. The limits are derived with all other coefficients set to zero.\label{limits_cg}}
\end{figure} 

 \begin{table}[hbt!]
\caption{Limits at 95\% C.L for the $\bar{c}_{\gamma}$, $\tilde{c}_{\gamma}$, $\bar{c}_{g}$ and $\tilde{c}_{g}$ Wilson coefficients considering  $\delta_{sys}$=0, 0.5\% and 2\% of systematic errors with an integrated luminosity of 30 ab$^{-1}$. }  
\begin{ruledtabular}\label{limit}
\begin{tabular}{llcc}
Coefficient&$\delta_{sys}$ &95\% C.L.  Limits  \\\hline
                                &0&$[-1.21; 1.21]\times10^{-6}$ $\cup$ $[4.20; 4.20]\times10^{-3}$  \\
$\bar{c}_{\gamma}$&$0.5\%$&$[-7.56; 7.84]\times10^{-5}$ $\cup$ $[4.12; 4.28]\times10^{-3}$    \\
                                &$2 \%$&$[-2.88; 3.35]\times10^{-4}$ $\cup$ $[3.87; 4.49]\times10^{-3}$ \\ ~\\
                                &0&$[-6.96; 6.96]\times10^{-5}$   \\
$\tilde{c}_{\gamma}$&$0.5\%$&$[-5.56; 5.56]\times10^{-4}$    \\
                                &$2\%$&$[-1.11; 1.11]\times10^{-3}$    \\ ~\\                               
                                 &0&$[-4.88; -4.88]\times10^{-4}$ $\cup$ $[3.37; 3.61]\times10^{-6}$  \\
$\bar{c}_{g}$&$0.5\%$&$[-4.96; -4.80]\times10^{-4}$ $\cup$ $[-0.46; 1.13]\times10^{-5}$    \\
                                &$2\%$&$[-5.18; -4.54]\times10^{-4}$ $\cup$ $[-3.08; 3.35]\times10^{-5}$  \\ ~\\                                                  
                                &0&$[-4.06; 4.06]\times10^{-5}$  \\
$\tilde{c}_{g}$&$0.5\%$&$[-7.25; 7.25]\times10^{-5}$   \\
                                &$2\%$&$[-1.27; 1.27]\times10^{-4}$   \\
  \end{tabular}
\end{ruledtabular}
\end{table}

Fig.~\ref{limits_cg} shows the obtained results at 95\% C.L. from two-parameter analysis of the $\bar{c}_{g}$-$\tilde{c}_{g}$ (on the right) and $\bar{c}_{\gamma}$-$\tilde{c}_{\gamma}$ (on the left) couplings considering  $\delta_{sys}$=0, 0.5\% and 2\% of systematic errors at $L_{int}=1$  ab$^{-1}$ for 100 TeV center of mass energy. From these figures, the limits on dimension-6 Higgs-gauge boson couplings $\bar{c}_{\gamma} $ and $\tilde{c}_{\gamma}$ at 95\% C.L. without systematic error at $L_{int}=1$ ab$^{-1}$ are [-6.60; 6.62]$\times10^{-6}$  $\cup$ [4.20; 4.21]$\times10^{-3}$ and [-1.63; 1.63]$\times10^{-4}$, respectively while the limits on $\bar{c}_{g} $ and $\tilde{c}_{g}$ are [-4.89; -4.87]$\times10^{-4}$  $\cup$ [2.84; 4.13]$\times10^{-6}$ and [-4.34; 4.34]$\times10^{-5}$. 
ATLAS collaboration reported 95\% C.L. limits on these couplings based on a fit to five differential cross sections with an integrated luminosity of 20.3 fb$^{-1}$ at $\sqrt s$=8 TeV in $H\to\gamma\gamma$ decay channel as [-7.4; 5.7]$\times10^{-4}$ $\cup$ [3.8; 5.1]$\times10^{-3}$ ([-0.7; 1.3]$\times10^{-4}$ $\cup$ [-5.8; -3.8]$\times10^{-4}$) and [-1.8; 1.8]$\times10^{-3}$ ([-2.4; 2.4]$\times10^{-4}$) for $\bar{c}_{\gamma} $ ($\bar{c}_{g} $)and $\tilde{c}_{\gamma}$ ( $\tilde{c}_{g}$), respectively  \cite{Aad:2015tna}. They also performed the similar analysis using $\sqrt s$ = 13 TeV data with $L_{int}$=36.1 fb$^{-1}$ and obtained limits on $\bar{c}_{g} $ and $\tilde{c}_{g}$ are  [-0.8; 0.1]$\times10^{-4}$ $\cup$ [-4.6; -3.8]$\times10^{-4}$ and [-1.0; 0.9]$\times10^{-4}$ while they did not consider $\bar{c}_{\gamma} $ and $\tilde{c}_{\gamma}$ couplings due to the lack of sensitivity of the $H\to\gamma\gamma$ decay channel \cite{Aaboud:2018xdt}. Results of follow up study with increase luminosity ($L_{int}$=139 fb$^{-1}$) at $\sqrt s$=13 TeV by ATLAS collaboration are [-1.1; 1.1]$\times10^{-4}$ ([-0.26; 0.26]$\times10^{-4}$ ) and [-2.8; 4.3]$\times10^{-4}$ ([-1.3; 1.1]$\times10^{-4}$) for $\bar{c}_{\gamma} $ ($\bar{c}_{g} $)and $\tilde{c}_{\gamma}$ ( $\tilde{c}_{g}$), respectively \cite{ATLAS:2019jst}. Our obtained results including detector effects for 100~TeV center of mass energy with an integrated luminosity of 30 ab$^{-1}$ with and without systematic errors are given in Table \ref{limit}. We report at least one order or more better than current experimental limits reported by ATLAS experiment. More specifically we obtained $[-1.21; 1.21]\times10^{-6}$ $\cup$ $[4.20; 4.20]\times10^{-3}$ while ATLAS collaboration found [-1.1;1.1]$\times10^{-4}$ for $\bar{c}_{\gamma}$ coupling. Our limits on $\bar{c}_{g}$, $\tilde{c}_{g}$, $\bar{c}_{\gamma}$ and $\tilde{c}_{\gamma}$  couplings can be effected by systematic uncertainties. One can make several assumptions on the evolution of sources of uncertainties when presenting a realistic physics potential of FCC-hh for the process $pp\to\gamma\gamma$+n-jet. Among different possible scenarios considered in the literature \cite{Mangano:2020sao}, we discuss our results based on two of these: target detector performance (optimistic) and intermediate detector performance (realistic). Considering the time scale of the FCC, overall uncertainty on the Higgs production cross section may vary between 0.5\% and 1\% with improvements on the theoretical predictions for optimistic and realistic scenario, respectively. Systematic uncertainties on the integrated luminosity will be at the same order of the LHC. However new techniques to extract the luminosity can improve this to an optimistic (realistic) scenario of 0.5\% (1\%). Finally, we consider the systematic uncertainty on the photon reconstruction of 0.5 \% and 1\% for the optimistic and realistic scenario, respectively.
As you can see from Table \ref{limit}, we estimate one order of magnitude better limits on the couplings considered in this study than the current experimental results for optimistic scenario ($\delta_{sys}$=0.5\% ) whereas comparable limits with realistic scenario ($\delta_{sys}$=2\%). On the other hand, including other decay channels of the Higgs boson as in Refs. \cite{Aad:2015tna, Aaboud:2018xdt,ATLAS:2019jst},  one can obtain improved bounds on $\bar{c}_{g}$, $\tilde{c}_{g}$, $\bar{c}_{\gamma}$ and $\tilde{c}_{\gamma}$  couplings. 
\section{Conclusions}
Since $pp\to\gamma\gamma$+n-jet signal process is sensitive to the CP-conserving and CP-violating dimension-six operators of Higgs-gauge boson couplings in the gauge basis, we have investigated this process to determine the sensitivity interval of $\bar{c}_{\gamma} $, $\tilde{c}_{\gamma}$, $\bar{c}_{g} $ and $\tilde{c}_{g}$ couplings using leading-order strongly interacting light Higgs basis effective Lagrangian approach at FCC-hh ($\sqrt s$=100 TeV, L$_{int}$=1-30 ab$^{-1}$). Realistic detector effects are included in the analysis via Delphes card prepared for FCC-hh. Kinematic variables of both leading photons and diphoton system are investigated to find optimum cuts to obtain best limits on the couplings. We have found 95\% C.L. constraints on four Wilson coefficients by using transverse momentum distributions of diphoton system of signal process and the relevant SM backgrounds. Our results demonstrate that FCC-hh with $\sqrt s=100$ TeV and $L_{int}$=30 ab$^{-1}$ will be able to obtain best limits on $\bar{c}_{\gamma} $ and $\tilde{c}_{\gamma}$ ($\bar{c}_{g} $ and $\tilde{c}_{g}$) couplings as $[-1.21; 1.21]\times10^{-6}$ $\cup$ $[4.20; 4.20]\times10^{-3}$ and $[-6.96; 6.96]\times10^{-5}$ ($[-4.88; -4.88]\times10^{-4}$ $\cup$ $[3.37; 3.61]\times10^{-6}$ and $[-4.06; 4.06]\times10^{-5}$) without systematic errors, respectively. The result of this study also shows that finding lower bounds would benefit from the high luminosity when the systematic uncertainties are well reduced below 2\% for FCC-hh. 
  
\begin{acknowledgments}
Authors would like to acknowledge with gratitude the partial support by Turkish Atomic Energy Authority (TAEK) under the grant No. 2018TAEK(CERN)A5.H6.F2-20. Authors also would like to thank the theory division of CERN where this work was initiated for the hospitality.
\end{acknowledgments}

\end{document}